# Variability Morphologies In The Color-Magnitude Diagram Searching For Secular Variability

Maxime Spano, Nami Mowlavi, Laurent Eyer, Gilbert Burki

*Geneva Observatory, University of Geneva, 51 Chemin des Maillettes, CH-1290 Sauverny, Switzerland*

**Abstract**. This work is part of an effort to detect secular variable objects in large scale surveys by analysing their path in color-magnitude diagrams. To this aim, we first present the variability morphologies in the V/V-I diagram of several types of variable stars. They comprise both periodic and non periodic variable stars from the Large Magellanic Cloud, such as classical Cepheids, long period variables or Be and R Coronae Borealis stars, as well as two of the detected secular variable stars in the Galaxy, FG Sge and V4334 Sgr. The study of the different variability morphologies allows the identification of regions in the color-magnitude diagram where those secular variable stars could be detected. We also estimate the number of such secular variable stars expected in the Large Magellanic Cloud.

## INTRODUCTION

At some evolutionary phases, the stellar evolution timescale becomes comparable to the human lifetime. The evolution of these stars, called secular variable stars, then becomes detectable in real time, providing a unique way to directly compare observations with model predictions of stellar evolution. Examples of secular variable stars include asymptotic and post-Asymptotic Giant Branch stars (AGB, post-AGB), luminous blue variables or planetary nebulae, to cite only a few.

In this work, we concentrate on the detection of post-AGB cases of secular variable stars in the field of the Large Magellanic Cloud (LMC), through the analysis of the OGLE database.

## FINDING SECULAR VARIABLE STARS

In order to find secular variable candidates in the Color-Magnitude Diagram (CMD), we need to disentangle them from the other variable stars. The first step is to understand the global structure of the CMD used in this study, illustrated by Fig. 1. The interested reader is refered to Alcock et al. (2000) for more explanations on the identified main structures A to G. As very few cases of secular variability are expected to be found, we need samples of stars as large as possible. Major surveys of our Galaxy and nearby galaxies are thus favored. OGLE-II and III surveys of the LMC, providing BVI photometry for resp. 7 and 35 millions stars (Udalski et al. 1997, Udalski et al. 2000, Szymański 2005, Udalski et al. 2008) have been chosen here. All the variable stars identified so far in the OGLE database are plotted in the V/V-I CMD shown in Fig. 2.

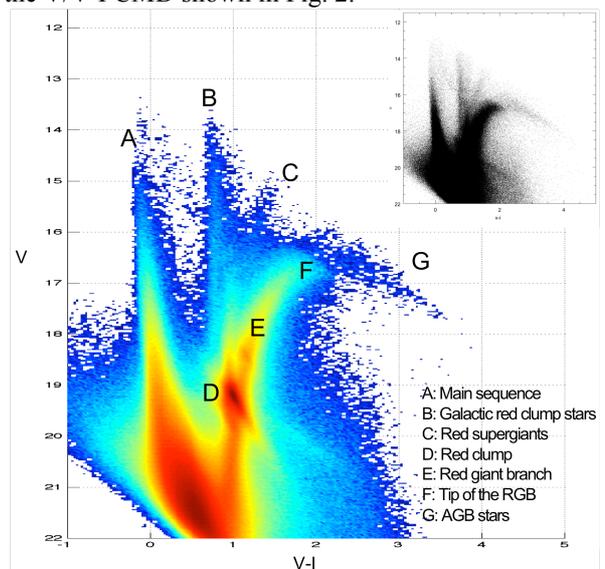

**FIGURE 1.** Hess CMD of 25% of the OGLE III Survey (subfields 1 and 4 selected among the 8 available for each of the 115 fields). The number density of stars increasing from blue to red.

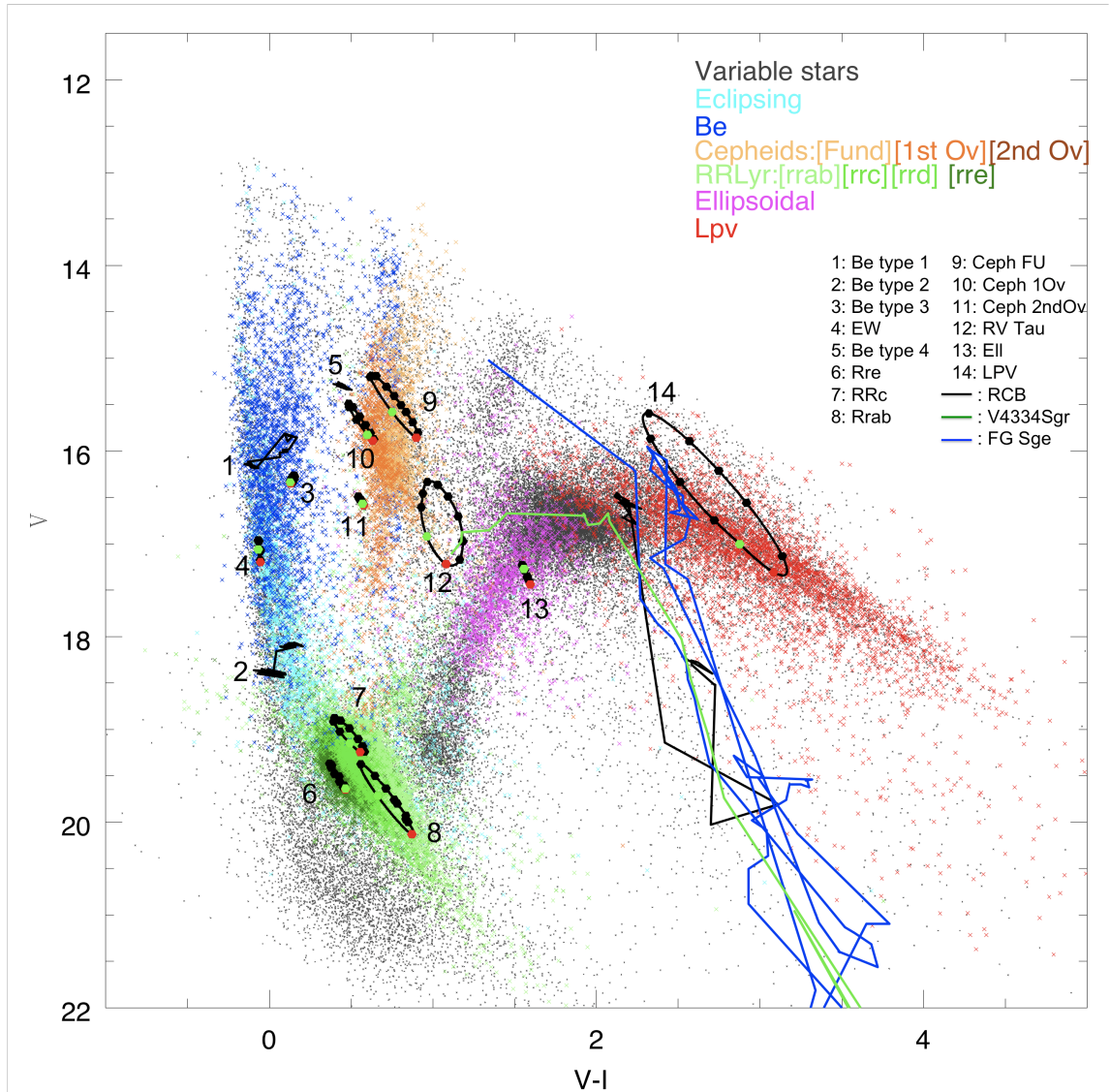

**FIGURE 2.** CMD of the LMC based on the variable stars taken from the OGLE II catalogue, except for RR Lyrae and Cepheids which are from OGLE III (Soszynski et al. 2009, Soszynski et al. 2008) and Be stars from Sabogal et al. 2005. The morphology of typical variables stars is represented with connected dots, equally spaced in phase, which inform on the speed of variation. The rotation sense of motion goes from the red to the green dot. For comparison V4334 Sgr and FG Sge are placed according to the estimated distance of 3.5 kpc of Miller Bertolami & Althaus (2007) and 3.1 kpc from Taranova & Shenavrin (2002), and assuming a distance modulus of 18.54 for the LMC (Keller & Wood, 2006). The motion of a RCB in the LMC is also shown. Figure available in color in the electronic version.

Fig. 2 allows to estimate the relative amplitudes of magnitude and color changes for the different types of variable stars, as well as the corresponding morphologies of their variability. The behavior of two secular variables (FG Sge and V4334 Sgr, rescaled at the distance of the LMC) is also shown in that figure.

## THE CASE OF POST-AGB STARS

Leaving the Thermally Pulsating AGB phase (TP-AGB), post-AGB stars cross the Hertzsprung-Russell (HR) diagram to end up their life as white dwarfs. Some of them, however, may experience a last thermal pulse on this path, depending on their interpulse and HR diagram crossing time scales. For these ones, this event leads to a drop followed by a recovery of luminosity together with a drop of Teff (cf. Fig. 14 of Blöcker 1995), on a time scale from about a year to a century, depending on when this last pulse occurs. In all cases it is expected to rapidly cross the HR diagram and result in a so called 'born-again' AGB star.

Few examples of secular variables have been observed, the best known cases are FG Sge, V4334 Sgr (Sakurai's object) and V605 Aql, but there is still no full agreement with models (see van Hoof et al. 2007 for example).

In Fig. 2 the tracks followed by Fg Sge and V4334 Sgr show a distinct behavior compared to the ones of the classical variables, until they reach the RCB phase. In Fig. 3, available B band data allow us to see earlier effects of the last thermal pulse, like the monotonic rise in V band at almost constant B-V or reddening at nearly constant V magnitude. Such a last thermal pulse occuring in the LMC would thus easily be spotable.

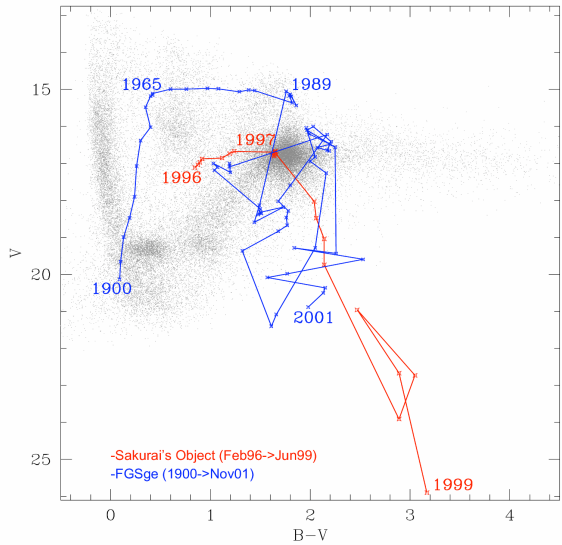

**FIGURE 3.** Observed evolution paths of FG Sge and V4334 Sgr (Sakurai's Object) on the LMC CMD.

## ESTIMATING THE NUMBER OF POST-AGB SECULAR VARIABLE STARS

A simple way to estimate the percentage of post-AGB stars that experience a last thermal pulse is to compare the duration of the post-AGB phase $\Delta t_{P-AGB}$ to the interpulse duration $\Delta t_{interpulse}$. According to the work of Vassiliadis & Wood (1994), at the corresponding LMC metallicity (Z=0.008), the ratio $R_{ltp} = \frac{\Delta t_{P-AGB}}{\Delta t_{interpulse}}$ ranges between 0.14 and 0.30 for post-AGB stars of initial masses between 1 and 2.5$M_\odot$ and metallicities between 0.004 and 0.016. On the other hand, the duration of the AGB phase is expected to be of the order of a thousand times longer than that of the post-AGB phase in this mass range at Z=0.008 (Karakas, Phd thesis 2003). We thus expect of the order of one post-AGB star out of ten thousand to experience a last thermal pulse. Two additional facts are favorable to our search. First, low mass stars are the most numerous component of galaxies. And second, they have shorter secular variation time scales after a last thermal pulse compared to more massive post-AGB stars (Schönberner 2008).

The number of AGB stars detected by OGLE-III is estimated to be ~20000 if we count all the objects falling in the 'G' branch of Fig. 1. Hence, possibly a few cases of secular variable post-AGB stars could be present in the LMC OGLE-III data base.

## CONCLUSIONS

The present analysis gives a positive hope on the possibility to find a secular variable post-AGB star in the OGLE data of the Magellanic Clouds. The variability morphology in the CMD helps to select areas where secular variable candidates can be noticed compare to the classical ones. The secular changes of post-AGB stars should be detectable thanks to the monotonic changes in their magnitude and color, as well as due to the fact that their motion in the CMD differs from that of other variable stars for timescale higher than a year or so. We are currently working on the identification of such candidates.

## ACKNOWLEDGMENTS

We thank the Swiss National Science Foundation for the continuous support to this project.

## REFERENCES

C. Alcock et al., AJ, 2000, 119, 2194.
V.P. Arkhipova et al., AstL, 2003, 29, 763.
T. Blöcker, A&A, 1995, 299, 755.
H.W. Duerbeck et al., AJ, 1997, 114, 1657.
H.W. Duerbeck et al., AJ, 2000, 119, 2360.
J.D. Fernie, ApJ, 1975, 200, 392.
A.M Van Genderen and A. Gautschy, A&A, 1995, 294, 453.
P.A.M. van Hoof et al., A&A, 2007, 471L, 9.
A. Karakas, "Asymptotic Giant Branch Stars: their influence on binary systems and the interstellar medium" Ph.D. Thesis, Monash University, 2003.
S.C. Keller and P.R. Wood, ApJ, 2006, 642, 834.
M.M. Miller Bertolami and L.G. Althaus, MNRAS, 2007, 380, 763.
B.E. Sabogal, R.E. Mennickent, G. Pietrzyński, W. Gieren, MNRAS, 2005, 361, 1055.
D. Schönberner, ASPC, 2008, 391, 139.
I. Soszynski et al., AcA, 2008, 58, 163.
I. Soszynski et al., AcA, 2009, 59, 1.
M.K. Szymański, AcA, 2005, 55, 43.
O.G. Taranova and V.I. Shenavrin, ARep, 2002, 46, 1010.
A. Udalski and M. Kubiak and M. Szymanski, AcA, 1997, 47, 319.
A. Udalski et al., AcA, 2000, 50, 307.
A. Udalski et al., AcA, 2008, 58, 89.
E. Vassiliadis and P.R. Wood, ApJS, 1994, 92, 125.